\documentclass{article}

\usepackage{microtype}
\usepackage{graphicx}
\usepackage{multirow}
\usepackage{booktabs}
\usepackage{subcaption}
\usepackage{verbatimbox}
\usepackage[table]{xcolor}
\usepackage{amssymb}
\usepackage{pifont}
\newcommand{\cmark}{\ding{51}}%
\newcommand{\xmark}{\ding{55}}%
\usepackage{bbm}

\usepackage{hyperref}

\usepackage[accepted]{icml2023}

\usepackage{amsmath}
\usepackage{amssymb}
\usepackage{mathtools}
\usepackage{amsthm}
\usepackage{comment}

\usepackage[capitalize,noabbrev]{cleveref}

\theoremstyle{plain}

\theoremstyle{definition}

\theoremstyle{remark}

\icmltitlerunning{Zero-Shot ECG Classification with Multimodal Learning and Test-time Clinical Knowledge Enhancement}

\begin{document}

\twocolumn[
\icmltitle{
Zero-Shot ECG Classification with Multimodal Learning and Test-time Clinical Knowledge Enhancement}



\icmlsetsymbol{equal}{*}

\begin{icmlauthorlist}
\icmlauthor{Che Liu}{icl}
\icmlauthor{Zhongwei Wan}{osu}
\icmlauthor{Cheng Ouyang}{icl}
\icmlauthor{Anand Shah}{icl}
\icmlauthor{Wenjia Bai}{icl}
\icmlauthor{Rossella Arcucci}{icl}
\end{icmlauthorlist}

\icmlaffiliation{icl}{Imperial College London}
\icmlaffiliation{osu}{Ohio State University}

\icmlcorrespondingauthor{Che Liu}{che.liu21@imperial.ac.uk}

\icmlkeywords{Machine Learning, ICML}

\vskip 0.3in
]



\printAffiliationsAndNotice{}  

\begin{abstract}
\small{
Electrocardiograms (ECGs) are non-invasive diagnostic tools crucial for detecting cardiac arrhythmic diseases in clinical practice. 
While ECG Self-supervised Learning (eSSL) methods show promise in representation learning from unannotated ECG data, they often overlook the clinical knowledge that can be found in reports. 
This oversight and the requirement for annotated samples for downstream tasks limit eSSL's versatility.
In this work, we address these issues with the \textbf{M}ultimodal \textbf{E}CG \textbf{R}epresentation \textbf{L}earning (\textbf{MERL}) framework. 
Through multimodal learning on ECG records and associated reports, MERL is capable of performing zero-shot ECG classification with text prompts, eliminating the need for training data in downstream tasks.
At test time, we propose the \textbf{C}linical \textbf{K}nowledge \textbf{E}nhanced \textbf{P}rompt \textbf{E}ngineering (\textbf{CKEPE}) approach, which uses Large Language Models (LLMs) to exploit external expert-verified clinical knowledge databases, generating more descriptive prompts and reducing hallucinations in LLM-generated content to boost zero-shot classification.
Based on MERL, we perform the first benchmark across six public ECG datasets, showing the superior performance of MERL compared against eSSL methods. 
Notably, MERL achieves an average AUC score of 75.2\% in zero-shot classification (\textbf{without training data}), 3.2\% higher than linear probed eSSL methods with 10\% annotated training data, averaged across all six datasets.}
\footnote{Code and models are available at \href{https://github.com/cheliu-computation/MERL}{https://github.com/cheliu-computation/MERL}}.

\end{abstract}

\section{Introduction}
\label{sec: intro}

\begin{figure}[t!]
    \centering
    \includegraphics[width=0.99\linewidth]{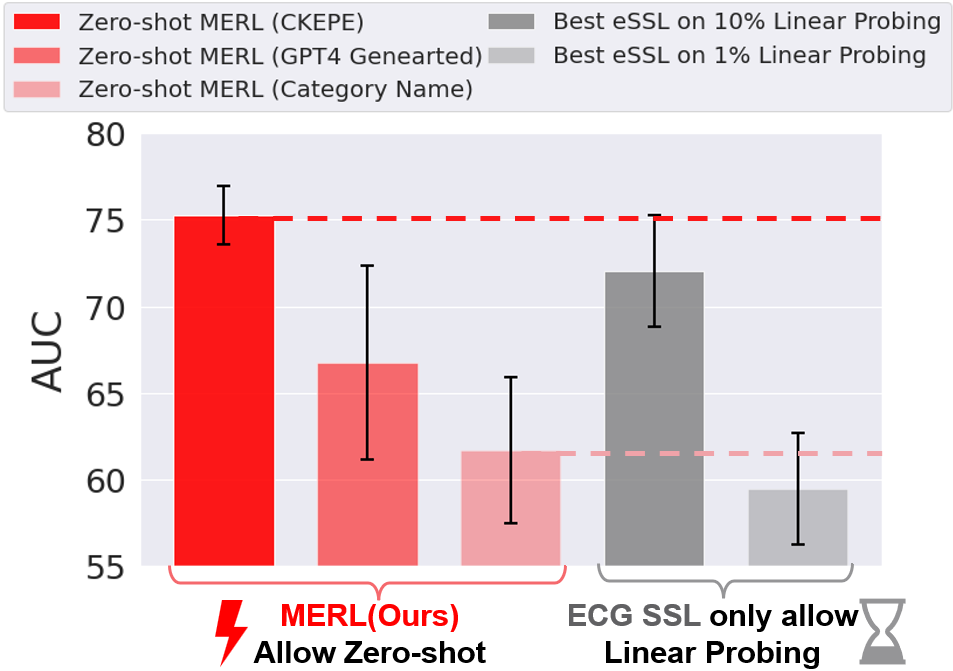}
    \vspace{-10pt}
    \caption{
    We demonstrate \textcolor{red}{MERL}, even without training samples and prompt engineering, surpasses the best-performing eSSL with 1\% data linear probing from Tab.~\ref{tab:linear-cls}. Additionally, zero-shot \textcolor{red}{MERL} enhanced with our CKEPE outperforms the best eSSL results obtained from 10\% data linear probing.
    }
    \label{fig: linear vs zero}
    \vspace{-15pt}
\end{figure}

Supervised learning methods effectively classify cardiac conditions using Electrocardiogram (ECG), a common clinical data for monitoring heart electrical activity~\cite{scdnn,SPN,SPNv2}.
However, these methods require large-scale data with high-quality annotations and expert review.
To reduce dependence on annotations, ECG self-supervised learning (eSSL) has made significant strides by utilizing the rich resource of unlabeled ECG records. 
Current eSSL techniques predominantly fall into two categories: contrastive and generative~\cite{tstcc,clocs,astcl,crt,stmem}. 
Contrastive eSSL (C-eSSL) focuses on learning discriminative ECG features by differentiating between augmented positive and negative samples~\cite{simclr,mocov3,astcl,tstcc,clocs,astcl}, whereas generative eSSL (G-eSSL) aims at reconstructing original signals from their masked versions~\cite{stmem,crt}. 

\begin{figure*}[t!]
    \centering
    \includegraphics[width=0.99\linewidth]{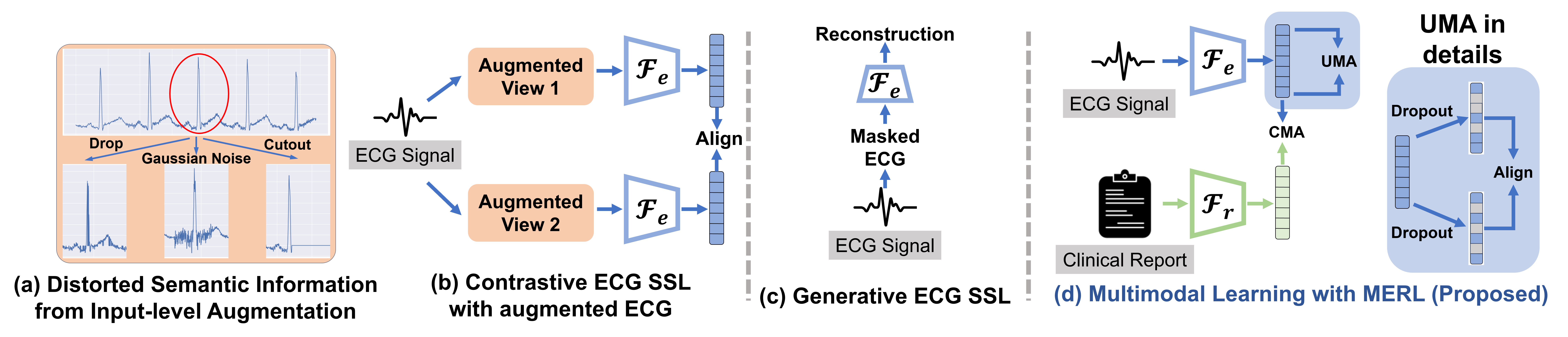}
    \vspace{-7pt}
    \caption{\textbf{(a)} Commonly used naive input-level data augmentation distorts semantics of ECG records, leading to sub-optimal representation learning performance. \textbf{(b)} Illustration of existing eSSL approaches. Their contrastive learning framework necessitates these naively augmented ECG signals. \textbf{(c)} Existing generative eSSL employs signal reconstruction as self-supervision task while being agnostic to the semantic meaning of ECG. \textbf{(d)} The proposed \textbf{MERL}, designed for multimodal ECG learning, leverages both ECG records and clinical reports for representation learning through \textbf{C}ross-\textbf{M}odal \textbf{A}lignment (\textbf{CMA}). MERL addresses the drawbacks of naive input-level augmentation by opting for latent augmentation (dropout) to prevent pattern corruption, and it enhances ECG learning through \textbf{U}ni-\textbf{M}odal \textbf{A}lignment (\textbf{UMA}). $\mathcal{F}_{e}$ denotes the ECG encoder, and $\mathcal{F}_r$ represents the text report encoder.}
    \label{fig: frame}
    \vspace{-5mm}
\end{figure*}

These methods, however, encounter two primary challenges:
\noindent\textbf{Semantic Distortion from Input-Level Augmentation in C-eSSL.}
Recent developments in C-eSSL for ECG representation learning often create two augmented views from the \textit{same} ECG signal to build \textit{positive} pairs (\textit{i.e.}, aligning their features to be identical) and consider views from different ECG records as \textit{negative} pairs (\textit{i.e.}, with differing features) within a contrastive framework \cite{simclr,mocov3,astcl,tstcc,clocs,astcl}. 
However, current ECG augmentation strategies, such as cutout and drop \cite{stmem}, could distort semantic information in ECG signals \cite{badpair}, as shown in Fig. \ref{fig: frame} (a). 
Consequently, the use of ECG with distorted semantics in positive and negative pairs compromises the quality of ECG representations learned through C-eSSL approaches.

\noindent\textbf{Limited High-level Semantics in G-eSSL.}
As shown Fig. \ref{fig: frame} (c), G-eSSL methods \cite{crt,stmem} learn to restore low-level signal patterns (e.g., local signal intensities and waveforms) while overlooking high-level semantics such as the diseases behind \cite{liu2023improving,liu2023pixmim,he2022masked}.
However, high-level semantics are essential for downstream ECG classification tasks. 
Therefore, the lack of high-level semantics in ECG representation can limit the performance of pre-trained models in these tasks.

Besides issues with distorted semantic information and missing high-level semantics, eSSL approaches are incapable of zero-shot classification as they only focus on extracting signal patterns, agnostic to the clinical concepts behind, limiting their versatility and risking distribution shifts in downstream tasks.

Multimodal learning has emerged as a promising approach for learning high-level semantics from other modalities, such as clinical reports \cite{radford2021learning,liu2023utilizing,liu2023imitate,wan2023med}. 
This strategy has achieved significant progress in the field of medical imaging and radiology reports \cite{liu2023m,chen2023generative,liu2023g2d,liu2023t3d}. 
However, as these approaches are mostly designed for image-language domains, their efficacy in ECG and their associated reports remains underexplored. 
The inherent differences between data modalities (signals vs. images) and the distinct nature of ECG report versus radiology reports pose challenges in directly applying existing multimodal methods from radiography to ECG records.
Furthermore, the text prompt for zero-shot classification, a new capability enabled by multimodal learning, requires a dynamic approach to generate more descriptive prompts at test time. It can also leverage external knowledge databases verified by clinical experts to ensure the quality and reliability of the generated prompts, moving beyond crude category names or fixed templates.
Additionally, benchmarks are needed to thoroughly assess the influence of large-scale data on ECG representation learning and evaluate the performance and robustness of these methods across a range of cardiac conditions in public datasets.

To address these challenges, this work has four contributions:
\begin{itemize}

\item  We propose a straightforward yet effective \textbf{M}ultimodal \textbf{E}CG \textbf{R}epresentation \textbf{L}earning framework (MERL) for ECG signals and associated reports. 
Unlike eSSL, MERL is capable of \textit{zero-shot} classification. 
Zero-shot MERL even outperforms linear probed eSSL with 10\% data, as averaged across six datasets.
Furthermore, linear probed MERL outperform eSSL across all downstream datasets and data ratios.

\item At training time, we introduce \textbf{C}ross-\textbf{M}odal \textbf{A}lignment (\textbf{CMA}) and \textbf{U}ni-\textbf{M}odal \textbf{A}lignment (\textbf{UMA}) for multimodal representation learning with ECG records and paired clinical reports, with augmentation at the latent level rather than the naive signal level to avoid semantic distortion.


\item At test time, we design \textbf{C}linical \textbf{K}nowledge \textbf{E}nhanced \textbf{P}rompt \textbf{E}ngineering (\textbf{CKEPE}), utilizing LLMs to dynamically generate customized prompts for zero-shot classification by extracting and restructuring knowledge from customer-provided knowledge databases verified by clinical experts.

\item To facilitate future research, we build the first benchmark by pre-training MERL and 10 eSSL methods on the largest publicly available ECG dataset, evaluating their performance on six diverse datasets covering over 100 cardiac conditions. This benchmark, covering zero-shot, linear probing, and data distribution transfer scenarios, assesses the quality and robustness of learned ECG representations.

\end{itemize}

\section{Related Work}
\label{sec: realted}

\subsection{Representation Learning with Multimodal Medical Data}
Various studies have explored medical multimodal learning, but mostly in radiography \cite{liu2023m,liu2023imitate,wan2023med,liu2023utilizing,liu2023g2d,liu2023t3d,chen2023generative}, with a focus on aligning global and local image features with radiology reports. 
Compared with images, ECG signals pose a unique challenge due to its global temporal and spatial structures, which span the entire prolonged signal period and are difficult to characterize at a local level.
\cite{lalam2023ecg} demonstrated the effectiveness of ECG and EHR multimodal learning, although their approach was limited to a private dataset. 
\cite{li2023frozen,liu2023etp} attempt multimodal ECG learning for zero-shot classification but fall short: 
Their methods, crudely aligning signals with text, overlook distinct signal patterns, and their reliance on simple cardiac condition names as prompts misses critical clinical attributes, leading to sub-optimal performance. 
Furthermore, their limited evaluations on small datasets is inadequate for assessing the potential of multimodal ECG learning in complicated real-world scenarios.

\subsection{Self-supervised Learning for ECG on Signal Domain}
Recently, ECG self-supervised learning (eSSL) has proven beneficial for learning transferrable representations directly from unannotated ECG signals \cite{lai2023practical, simclr}. 
Among them contrastive eSSL methods such as CLOCS \cite{clocs} and ASTCL \cite{astcl} have advanced eSSL by exploring temporal and spatial invariance and employing adversarial learning, respectively. 
Generative eSSL techniques, as discussed in \cite{zhang2022maefe,sawano2022masked,stmem}, learn ECG representations through pretext tasks involving masked segment reconstruction. However, they face challenges in attaining high-level semantic representations.
Both contrastive and generative eSSL methods are often agnostic of high-level clinical domain knowledge. Therefore, there is still an unmet need for an effective unsupervised approach for learning semantically rich, transferable ECG representations.

\subsection{Customizing Prompt for Zero-shot Classification}
In zero-shot classification, the prompt's quality, often limited in traditional methods that use basic category names, is crucial for effective performance and classification \cite{radford2021learning}.
To improve this, \cite{menon2022visual, pratt2023does} uses LLMs to generate attribute-rich prompts, boosting performance. 
Yet, in medical field, where terminologies are highly specialized, prompts generated by non-specialist LLMs might be inaccurate or untrustable, leading to performance degradation and safety concerns. 
Instead of relying on limited knowledge encoded in non-specialized LLMs, we re-purpose LLMs for extracting and re-formatting specialized clinical knowledge from trustable, external sources such as online and local clinical knowledge databases. 
By this mean, we can efficiently create clinically relevant, structured prompts without additional annotations.

\section{Method}
\label{sec: method}

\subsection{Overview}
Our MERL framework learns transferable ECG representations directly from ECG signals and associated text reports. These learned representation can be then directly applied for zero-shot classification of unseen diseases. To achieve this, our framework is a synergy of a train-time ECG-report multimodal representation learning strategy and a test-time clinical knowledge enhanced prompt engineering approach.
Specifically, as depicted in Fig. \ref{fig: frame} (d), the representation learning strategy comprises \textbf{C}ross-\textbf{M}odal \textbf{A}lignment (\textbf{CMA}), detailed in Sec. \ref{sec: cma}, and \textbf{U}ni-Modal \textbf{A}lignment (\textbf{UMA}), described in Sec. \ref{sec: uma}. Additionally, \textbf{C}linical \textbf{K}nowledge \textbf{E}nhanced \textbf{P}rompt \textbf{E}ngineering (\textbf{CKEPE}) is introduced in Sec. \ref{sec: zeroshot pipeline}.


\subsection{Cross-Modal Alignment}
\label{sec: cma}
The \textbf{C}ross-\textbf{M}odal \textbf{A}lignment (\textbf{CMA}) aims to learn ECG features informed with clinical knowledge under report supervision. 
Specifically, given a training dataset $\mathcal{X}$ consisting of \(N\) ECG-report pairs, we represent each pair as \((\mathbf{e}_{i}, \mathbf{r}_{i})\), where \(\mathbf{e}_{i} \in \mathcal{E}\) denotes the raw ECG records and \(\mathbf{r}_{i} \in \mathcal{R}\) denotes the associated text report, respectively, with \(i = 1,2,3, ..., N\). 
In this framework, as shown in Fig. \ref{fig: frame} (d), two distinct encoders for ECG signals and text reports, symbolized as \( \mathcal{F}_{e} \) and \( \mathcal{F}_{r} \) respectively, transform the sample pair \(( \mathbf{e}_{i}, \mathbf{r}_{i} ) \) into the latent embedding space, represented as \( (\mathbf{z}_{e,i},  \mathbf{z}_{r,i} ) \).
Then dataset at feature-level is then denoted as \(\mathcal{X}=\left\{\left(\mathbf{z}_{e,1}, \mathbf{z}_{r,1}\right),\left(\mathbf{z}_{e,2}, \mathbf{z}_{r,2}\right), \ldots, \left(\mathbf{z}_{e,N}, \mathbf{z}_{r,N}\right)\right\}\), where \(\mathbf{z}_{e,i} = \mathcal{F}_{e}(\mathbf{e}_{i})\) and \(\mathbf{z}_{r,i} = \mathcal{F}_{r}(\mathbf{z}_{r,i})\).
After that, two non-linear projectors for ECG and text embedding, denoted as \( \mathcal{P}_{e} \) and \( \mathcal{P}_{r} \) respectively, convert \( \mathbf{z}_{e,i} \) and \( \mathbf{z}_{r,i} \) into the same dimensionality \( d \),  with \(\mathbf{\hat{z}}_{e,i} = \mathcal{P}_{e}(\mathbf{z}_{e,i})\), \(\mathbf{\hat{z}}_{r,i} = \mathcal{P}_{r}(\mathbf{z}_{r,i})\). 
Then, we compute the cosine similarities as \(s_{i,i}^{e2r} = \mathbf{\hat{e}}_{i}^{\top} \mathbf{\hat{r}}_{i}\) and \(s_{i,i}^{r2e} = \mathbf{\hat{r}}_{i}^{\top} \mathbf{\hat{e}}_{i}\), representing the ECG-report and report-ECG similarities, respectively. 
The loss function, \(\mathcal{L}_{\mathrm{CMA}}\), is then expressed as:

\begin{align}
\mathcal{L}^{e2r}_{i,j} = -\log \frac{\exp(s_{i,j}^{e2r} / \tau_{})}{\sum_{k=1}^{L}{\mathbbm{1}_{[k \neq i]}\exp(s_{i,k}^{e2r} / \tau_{})}}, \\
\mathcal{L}^{r2e}_{i,j} = -\log \frac{\exp(s_{i,j}^{r2e} / \tau_{})}{\sum_{k=1}^{L}{\mathbbm{1}_{[k \neq i]}\exp(s_{i,k}^{e2r} / \tau_{})}}, \\
\mathcal{L}_{\mathrm{CMA}}=\frac{1}{2 L} \sum_{i=1}^N\sum_{j=1}^N\left(\mathcal{L}^{e2r}_{i,j}+\mathcal{L}^{r2e}_{i,j}\right). 
\end{align}

Here, \(\mathcal{L}^{e2r}_{i,j}\) and \(\mathcal{L}^{r2e}_{i,j}\) represent the ECG-report and report-ECG cross-modal contrastive losses, respectively. The temperature hyper-parameter, denoted as \(\tau_{}\), is set to 0.07 in our study. Additionally, \(L\) signifies the batch size per step, being a subset of \(N\).

\subsection{Uni-Modal Alignment}
\label{sec: uma}
On top of CMA, we further employ \textbf{U}ni-\textbf{M}odal \textbf{A}lignment (\textbf{UMA}) to facilitate representation learning. UMA is formulated as contrastive learning operating on the signal domain only. 
To circumvent the distortion of semantic information caused by naive input-level data augmentation (shown in Fig. \ref{fig: frame} (a)), we use our proposed latent augmentation on the ECG embedding $\mathbf{z}_{e,i}$ to construct positive pairs for contrastive learning, which is illustrated in the blue shaded block of in Fig. \ref{fig: frame} (d).
Inspired by \cite{gao2021simcse}, to generate the positive pair $(\mathbf{z}_{e,i}^{1}, \mathbf{z}_{e,i}^{2})$, we adopt two independent dropout operations on the ECG embeddings separately.
Then, we use standard contrastive loss on the positive pair and treat other unpaired combinations as negative pairs.
The loss function of $\mathcal{L}_{\mathrm{UMA}}$ can be denoted as:
\begin{equation}
\begin{aligned}
\mathcal{L}_{\mathrm{UMA}} = -\frac{1}{L} \sum_{i=1}^N\sum_{j=1}^N&\log \frac{\mathrm{exp}(s_{i,j} / \tau)}{\sum_{k=1}^{L}{\mathbbm{1}_{[k \neq i]}\exp(s_{i,j} / \tau)}}, \\
\text{where\hspace{1mm}} s_{i,i}  &= \mathbf{z}_{e,i}^{1 \top}  \mathbf{z}_{e,i}^{2}, \\ 
\mathbf{z}_{e,i}^{1 }=\mathbf{z}_{e,i}^{}\odot &M^{1}, \hspace{1mm}M^{1}\sim \text { Bernoulli }(p),\\
\mathbf{z}_{e,i}^{2 }=\mathbf{z}_{e,i}^{}\odot &M^{2}, \hspace{1mm}M^{2}\sim \text { Bernoulli }(p).\\
\end{aligned}
\end{equation}
$M^{1}$ and $M^{2}$ represent the dropout masks, which have the same sizes as $\mathbf{z}_{e,i}$'s, with each entry independently sampled with dropout ratio $p$, which is set to 0.1. $\odot$ denotes element-wise multiplication. The ablation study for $p$ is detailed in Tab \ref{tab: abla drop ratio}.
Importantly, as the two dropout operations are independent, the ECG embeddings post-dropout will not be identical, thus avoiding a trivial solution. 

In summary, our model learns representative ECG features by jointly minimizing $\mathcal{L}_{\mathrm{UMA}}$ and $\mathcal{L}_{\mathrm{CMA}}$, 
and the overall training loss can be written as:
\begin{equation}
    \mathcal{L}_{\mathrm{total}} = \mathcal{L}_{\mathrm{CMA}} +   \mathcal{L}_{\mathrm{UMA}}\text{,}
\end{equation}


\subsection{Enhancing Zero-Shot Prompts with External Clinical Knowledge Databases}
\label{sec: zeroshot pipeline}
\begin{figure}[t!]
    \centering
    \includegraphics[width=0.99\linewidth]{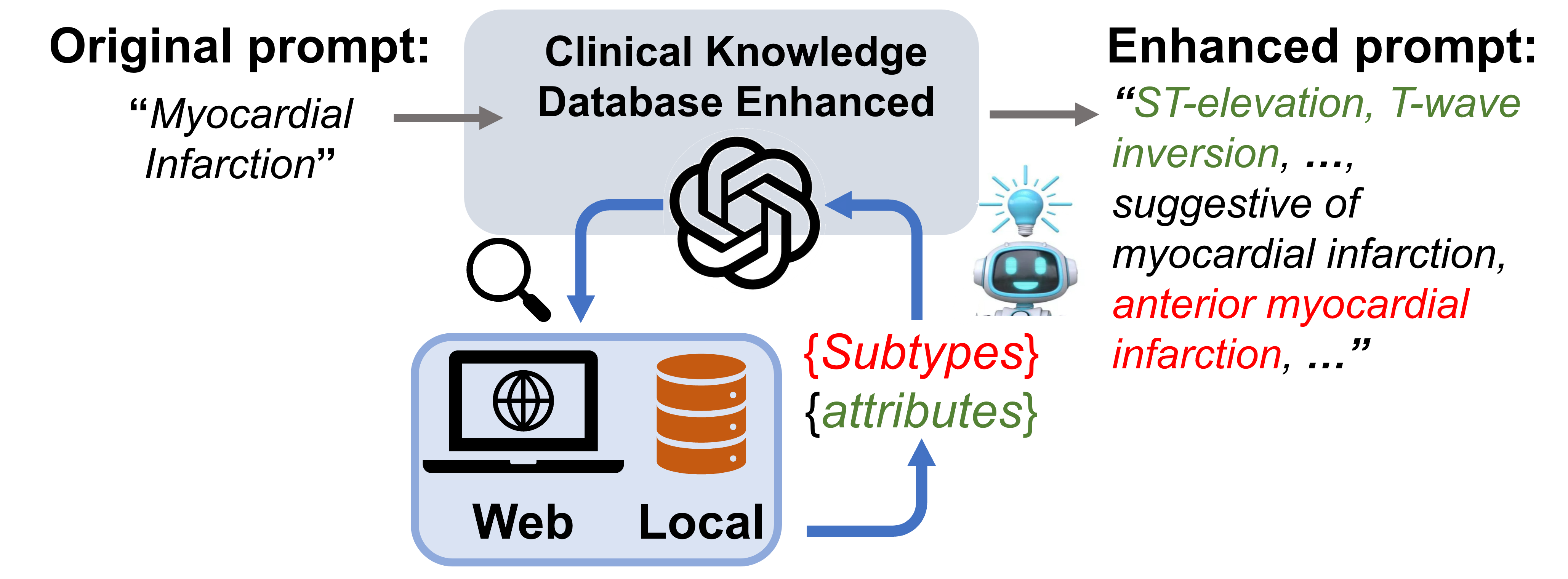}
    \vspace{-7pt}
    \caption{
At test time, we design the CKEPE pipeline for generating more descriptive prompts via LLM for zero-shot classification. In particular, we leverage the capability of LLM to extract clinical knowledge from trustworthy external knowledge databases verified by clinicians, then restructure this knowledge (e.g., subtypes or attributes of cardiac conditions) for prompt generation, with less hallucination from LLM.
}
    \label{fig: ckepe}
    \vspace{-5mm}
\end{figure}

The quality of text prompts, as part of input to a multimodality model, have been found to have significant impact on the zero-shot classification performance \cite{menon2022visual,pratt2023does,maniparambil2023enhancing}. 
The conventional method \cite{clip,convirt}  merely use names of cardiac conditions or a fixed template as text prompts for zero-shot classification. 
This approach, however, often perform poorly due to a lack of sufficient attributes (\textit{e.g.,} detailed description of signal patterns) and/or possible sub-categories at a finer level (\textit{e.g.,} possible subtypes of a clinical condition) for distinguishing the target semantic class.
While a powerful LLM can generate possible relevant attributes and subtypes for these categories as text prompts \cite{pratt2023does}, it incurs the risk of factual error/hallucination, often due to a lack of knowledge for the specialized domain \cite{liu2023exploring,hyland2023maira,umapathi2023med}. This risk renders this naive LLM-based paradigm unacceptable for medical applications.
To address this, we introduce \textbf{C}linical \textbf{K}nowledge \textbf{E}nhanced \textbf{P}rompt \textbf{E}ngineering (\textbf{CKEPE}).
Instead of directly, sourcing specialized knowledge from (possibly non-specialist) LLM, we leverage LLM to query and extract clinical knowledge from trustable, external knowledge bases, and reformat extracted knowledge as structured prompts. These clinical-knowledge-informed prompts contains descriptive attributes and possible finer-level labels (disease subtypes) of the targeted semantic class.

\noindent\textbf{Web and Local Clinical Knowledge Databases.}
To implement CKEPE, we initially prepare two databases rich in precise, expert-evaluated clinical knowledge. The first is the Systemized Nomenclature of Medicine – Clinical Terms (SNOMEDCT) \footnote{https://bioportal.bioontology.org/ontologies/SNOMEDCT}, a comprehensive, web-based database internationally validated for recording clinical information in structured clinical vocabulary \cite{stearns2001snomed}.

The second database is focused on Standard Communications Protocol (SCP) statements for ECG \cite{rubel2016scp}, which describes ECG states. As there is no publicly accessible database encompassing the entire SCP statement, we have constructed a local database by collecting relevant trustable sources from the internet \footnote{This database will be released after acceptance}.

\noindent\textbf{Searching, Thinking, and Generation.}
After preparing the web and local databases, we initially query GPT-4, the only current LLM featuring original web browsing capabilities, with: `\textit{Which attributes and subtypes does \texttt{<cardiac condition>} have? If this condition specifically describes symptoms or a subtype, please refrain from answering; otherwise, generate all possible scenarios.}' Following GPT-4's response, we enable its web browsing function towards the designated web database and we upload the local database file. We then instruct GPT-4 to search for relevant terms in both the web and local databases, ensuring the generated results exist in the clinical knowledge database and are relevant to the provided cardiac condition. Terms that are not found in either database are discarded. This verification step is automated by GPT-4 thinking.

Subsequently, we ask GPT-4 to reformulate the remaining terms into a structured text prompt for zero-shot classification, stylized like an ECG statement containing possible \textit{diseases subtypes} and descriptive \textit{attributes} describing signal patterns. 
As shown in Fig. \ref{fig: zeroshot res}, our prompts, unlike the original ones with only category names and fixed templates, are dynamic and tailored to specific cardiac conditions without the need for handcrafting. 
Moreover, the reformulated prompts adhere to clinically structured expressions, as all terms are cross-verified with the web and local clinical knowledge databases.

\section{Experiments}
\label{sec: exp}

\subsection{Pre-training Configuration}

\noindent \textbf{MIMIC-ECG.} 
In our study, we pre-train the MERL framework on the MIMIC-ECG dataset \cite{mimicecg}. 
This dataset contains 800,035 paired samples from 161,352 unique subjects. 
Each sample is composed of a raw ECG signal and its associated report, with every ECG recording sampled at 500Hz for a duration of 10 seconds.
To prepare the pre-training dataset, we executed the following procedures: \textbf{(1)} Exclude samples with an empty report or reports containing fewer than three words. \textbf{(2)} Substitute `NaN' and `Inf' values in ECG recordings with the average of the six neighboring points.
After these curation steps, our tailored dataset for training MERL contains 771,693 samples. 
Each sample includes an ECG record and its corresponding report. 

\noindent\textbf{Implementation.}
In pre-training stage, we employ a random initialized 1D-ResNet18 as the ECG encoder. 
For text encoding, we employ Med-CPT \cite{jin2023medcpt} by default. 
The impact of various text encoders on downstream performance is discussed in Sec \ref{sec: ablation}.
We select the AdamW optimizer, setting a learning rate of $2\times10^{-4}$ and a weight decay of $1\times10^{-5}$. 
We pre-train MERL for 50 epochs, applying a cosine annealing scheduler for learning rate adjustments. 
We maintain a batch size of 512 per GPU, with all experiments conducted on eight NVIDIA A100-40GB GPUs.

\subsection{Downstream Tasks Configuration}
We evaluate our framework on both zero-shot classification and linear probing, on three widely-used public datasets listed as follows, covering over 100 cardiac conditions. 
The details of the data split are shown in \textit{Appendix}.

\noindent \textbf{PTBXL.} This dataset \cite{wagner2020ptb} encompasses 21,837 ECG signals that were accumulated from 18,885 patients. The collected data consists of 12-lead ECG, each sampled at a rate of 500 Hz with a duration of 10 seconds. Based on ECG annotation protocol, there are four subsets with multi-label classification tasks: \textbf{Superclass} (5 categories), \textbf{Subclass} (23 categories), \textbf{Form} (19 categories), and \textbf{Rhythm} (12 categories). Notably, these four subsets have different number of samples. We follow the official data split \cite{wagner2020ptb} for the train:val:test split. 

\noindent \textbf{CPSC2018.} This publicly accessible dataset \cite{cpsc2018} comprises 6,877 standard 12-lead ECG records, each sampled at a rate of 500 Hz, and the duration of these records ranges from 6 to 60 seconds. The dataset is annotated with 9 distinct labels. We split the dataset as 70\%:10\%:20\% for the train:val:test split.

\noindent \textbf{Chapman-Shaoxing-Ningbo (CSN).} This publicly accessible dataset \cite{csn2,csn1} comprises 45,152 standard 12-lead ECG records, each sampled at a rate of 500 Hz with a duration of 10 seconds. We drop the ECG signals with `unknown' annotation, the curated version dataset has 23,026 ECG records with 38 distinct labels. The dataset is split into 70\%:10\%:20\%.

\noindent\textbf{Implementation.}
For linear probing, we keep the ECG encoder frozen and update only the parameters of a newly initialized linear classifier. 
We conduct linear probing for each task, utilizing ${1\%, 10\%, 100\%}$ of the training data. 
These configurations are consistent across all linear probing classification tasks. 
For zero-shot classification, we freeze the whole model, and use CKEPE to customize the prompt for each category. 
We compute the similarity between the ECG embedding and prompt embedding as the classification probability for the category associated with the prompt.
For all downstream tasks, we use macro AUC as the metric.
Further details, including specifics of the implementation, are provided in \textit{Appendix}.



\subsection{Evaluation on zero-shot learning}

\begin{figure}[t!]
    \centering
    \includegraphics[width=0.99\linewidth]{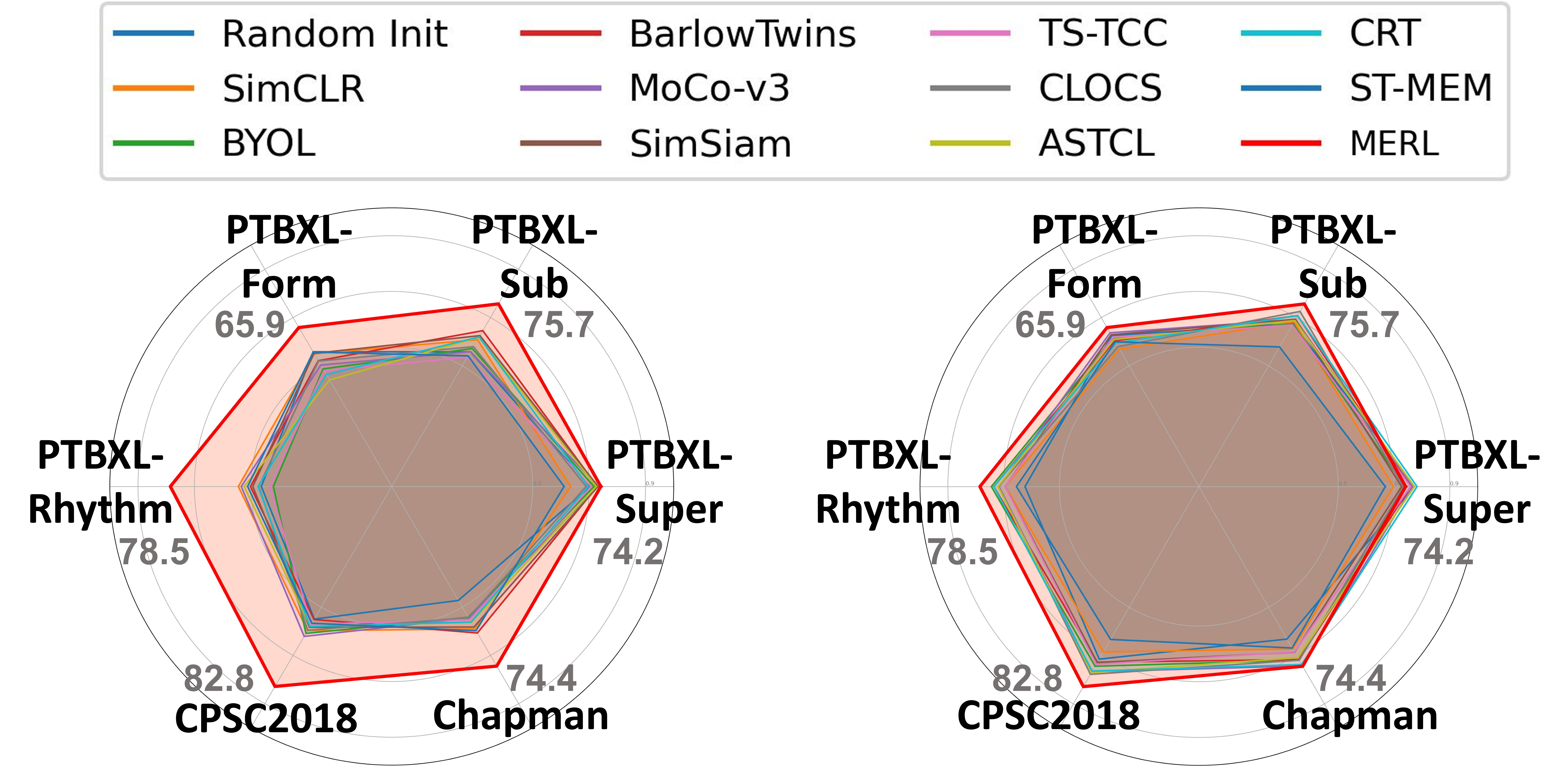}
    \vspace{-7pt}
    \caption{\textbf{Left}: Zero-shot \textcolor{red}{MERL} vs. linear probed eSSL with 1\% Data. \textbf{Right}: Zero-shot \textcolor{red}{MERL} vs. linear probed eSSL with 10\% Data. All performance are reported in the AUC score.}
    \label{fig: zeroshot res}
    \vspace{-7mm}
\end{figure}

Zero-shot classification using text prompts is a common task for assessing representation learning quality \cite{clip}. 
However, most research in ECG representation learning evaluates only linear probing. 
This limitation arises as these eSSL approaches merely operates on ECG signals only, without having a text encoder for receiving text prompts. 
We motivate zero-shot classification as a way to measure the quality and versatility of cross-modal ECG representations learned from clinical reports.

In our comprehensive analysis across six different datasets, we assess the performance of zero-shot MERL compared to eSSL approaches on linear probing, as shown in Fig. \ref{fig: linear vs zero}, \ref{fig: zeroshot res}. 
Fig. \ref{fig: linear vs zero} demonstrates how zero-shot MERL with three types of prompts outperforms eSSL methods that are linear probed with additional annotated data. 
Notably, zero-shot MERL, even without prompt enhancement, surpasses the top eSSL method linear probed with 1\% additional training data, achieving higher average AUC across six datasets. 
Furthermore, with CKEPE, zero-shot MERL exceeds the best eSSL method probed with additional 10\% data, underscoring the effectiveness of CKEPE and learned ECG representations from MERL.

We also present the performances of zero-shot MERL and eSSL performance on individual datasets, as shown on the left of Fig. \ref{fig: zeroshot res}.
Remarkably, zero-shot MERL demonstrates superior performance to eSSL with linear probing across all downstream datasets, even without additional training samples.
This underlines MERL's ability to learn robust, transferable cross-modal ECG features with clinically relevant knowledge from report supervision.

Interestingly, despite the significant overall performance gain, our method demonstrates a lesser advantage on the PTBXL-Super dataset.
This behavior may be attributed to the `simpler' nature of the PTBXL-Super dataset, which has only 5 broad categories (e.g., \textit{myocardial infarction}), compared to the 9–38 detailed categories (e.g., \textit{inferior myocardial infarction} or \textit{inferolateral myocardial infarction}) in other datasets.
Nevertheless, MERL notably outperforms other eSSL methods linear probed on 1\%-10\% additional annotated training data in the remaining five more challenging datasets.
These results demonstrate that clinical reports are a valuable supervision signal for ECG representation learning.

\subsection{Evaluation as ECG Representations}

\begin{table*}[ht]
\centering
\vspace{-7pt}
\caption{Linear probing results of MERL and eSSL methods. The best results are \textbf{bolded}, with \colorbox{gray!30}{gray} indicating the second highest.}
\scalebox{0.7}{    
\begin{tabular}{lccc|ccc|ccc|ccc|ccc|ccc}
    \toprule[1.2pt]
     & \multicolumn{3}{c}{PTBXL-Super} & \multicolumn{3}{c}{PTBXL-Sub} 
     & \multicolumn{3}{c}{PTBXL-Form} & \multicolumn{3}{c}{PTBXL-Rhythm} 
     & \multicolumn{3}{c}{CPSC2018} & \multicolumn{3}{c}{CSN} \\
    Method & 1\% & 10\% & 100\% & 1\% & 10\% & 100\% & 1\% & 10\% & 100\% 
           & 1\% & 10\% & 100\% & 1\% & 10\% & 100\% & 1\% & 10\% & 100\% \\
    \midrule[1.2pt]
    Random Init & 70.45 & 77.09 & \cellcolor{gray!30}81.61 & 55.82 & 67.60 & \cellcolor{gray!30}77.91 & \cellcolor{gray!30}55.82 & 62.54 & \cellcolor{gray!30}73.00 & 46.26 & 62.36 & \cellcolor{gray!30}79.29 & 54.96 & 71.47 & 78.33 & 47.22 & 63.17 & 73.13 \\
    SimCLR & 63.41 & 69.77 & 73.53 & 60.84 & 68.27 & 73.39 & 54.98 & 56.97 & 62.52 & 51.41 & 69.44 & 77.73 & 59.78 & 68.52 & 76.54 & 59.02 & 67.26 & 73.20 \\
    BYOL & 71.70 & 73.83 & 76.45 & 57.16 & 67.44 & 71.64 & 48.73 & 61.63 & 70.82 & 41.99 & \cellcolor{gray!30}74.40 & 77.17 & 60.88 & 74.42 & 78.75 & 54.20 & 71.92 & 74.69 \\
    BarlowTwins & 72.87 & 75.96 & 78.41 & \cellcolor{gray!30}62.57 & 70.84 & 74.34 & 52.12 & 60.39 & 66.14 & 50.12 & 73.54 & 77.62 & 55.12 & 72.75 & 78.39 & \cellcolor{gray!30}60.72 & 71.64 & 77.43 \\
    MoCo-v3 & \cellcolor{gray!30}73.19 & 76.65 & 78.26 & 55.88 & 69.21 & 76.69 & 50.32 & \cellcolor{gray!30}63.71 & 71.31 & 51.38 & 71.66 & 74.33 & \cellcolor{gray!30}62.13 & 76.74 & 75.29 & 54.61 & \cellcolor{gray!30}74.26 & 77.68 \\
    SimSiam & 73.15 & 72.70 & 75.63 & 62.52 & 69.31 & 76.38 & 55.16 & 62.91 & 71.31 & 49.30 & 69.47 & 75.92 & 58.35 & 72.89 & 75.31 & 58.25 & 68.61 & 77.41 \\
    TS-TCC & 70.73 & 75.88 & 78.91 & 53.54 & 66.98 & 77.87 & 48.04 & 61.79 & 71.18 & 43.34 & 69.48 & 78.23 & 57.07 & 73.62 & 78.72 & 55.26 & 68.48 & 76.79 \\
    CLOCS & 68.94 & 73.36 & 76.31 & 57.94 & \cellcolor{gray!30}72.55 & 76.24 & 51.97 & 57.96 & 72.65 & 47.19 & 71.88 & 76.31 & 59.59 & \cellcolor{gray!30}77.78 & 77.49 & 54.38 & 71.93 & 76.13 \\
    ASTCL & 72.51 & 77.31 & 81.02 & 61.86 & 68.77 & 76.51 & 44.14 & 60.93 & 66.99 & \cellcolor{gray!30}52.38 & 71.98 & 76.05 & 57.90 & 77.01 & 79.51 & 56.40 & 70.87 & 75.79 \\
    CRT & 69.68 & \cellcolor{gray!30}78.24 & 77.24 & 61.98 & 70.82 & 78.67 & 46.41 & 59.49 & 68.73 & 47.44 & 73.52 & 74.41 & 58.01 & 76.43 & \cellcolor{gray!30}82.03 & 56.21 & 73.70 & \cellcolor{gray!30}78.80 \\
    ST-MEM & 61.12 & 66.87 & 71.36 & 54.12 & 57.86 & 63.59 & 55.71 & 59.99 & 66.07 & 51.12 & 65.44 & 74.85 & 56.69 & 63.32 & 70.39 & 59.77 & 66.87 & 71.36 \\
    \midrule
    \textbf{MERL (Ours)} & \textbf{82.39} & \textbf{86.27} & \textbf{88.67} & \textbf{64.90} & \textbf{80.56} & \textbf{84.72} & \textbf{58.26} & \textbf{72.43} & \textbf{79.65} & \textbf{53.33} & \textbf{82.88} & \textbf{88.34} & \textbf{70.33} & \textbf{85.32} & \textbf{90.57} & \textbf{66.60} & \textbf{82.74} & \textbf{87.95} \\
    \bottomrule
\end{tabular}
}
\vspace{-3pt}
\label{tab:linear-cls}
\vspace{-2mm}
\end{table*}

\begin{table*}[ht]
\centering
\vspace{-7pt}
\caption{Results under data distribution shift: `Source Domain' denotes the dataset used for linear probing with the frozen pre-trained ECG encoder. `Target Domain' refers to the corresponding test set. We include only those target domain samples that match categories from the source domain. The top results are highlighted in bold, while the \colorbox{gray!30}{gray} color marks the second-highest achievements.}
    \scalebox{0.72}{    
    \begin{tabular}{l|c|c|cccccc}
        \toprule[1.2pt]
        \textbf{Source Domain} & \textbf{Zero-shot} & \textbf{Training Data} & \multicolumn{2}{c}{PTBXL-Super} & \multicolumn{2}{c}{CPSC2018} & \multicolumn{2}{c}{CSN} \\
        \textbf{Target Domain} & & \textbf{Ratio} & CPSC2018 & CSN & PTBXL-Super & CSN & PTBXL-Super & CPSC2018\\
        \midrule[1.2pt]
        Random Init & \xmark &  & 68.62 & 75.31 & 55.74 & 68.92 & 56.57 & 61.16 \\
        SimCLR~\cite{simclr} & \xmark &  & 69.62 & 73.05 & 56.65 & 66.36 & 59.74 & 62.11 \\
        BYOL~\cite{byol} & \xmark &  & 70.27 & 74.01 & 57.32 & 67.56 & 60.39 & 63.24\\
        BarlowTwins~\cite{zbontar2021barlow} & \xmark &  & 68.98 & 72.85 & 55.97 & 65.89 & 58.76 & 61.35 \\
        MoCo-v3~\cite{mocov3} & \xmark &  & 69.41 & 73.29 & 56.54 & 66.12 & 59.82 & 62.07\\
        SimSiam \cite{simsiam} & \xmark &  & 70.06 & 73.92 & 57.21 & 67.48 & 60.23 & 63.09\\
        TS-TCC \cite{tstcc} & \xmark & 100\% & 71.32 & 75.16 & 58.47 & 68.34 & 61.55 & 64.48\\
        CLOCS \cite{clocs} & \xmark &  & 68.79 & 72.64 & 55.86 & 65.73 & 58.69 & 61.27\\
        ASTCL \cite{astcl} & \xmark &  & 69.23 & 73.18 & 56.61 & 66.27 & 59.74 & 62.12\\
        CRT \cite{crt} & \xmark &  & 70.15 & 74.08 & 57.39 & 67.62 & 60.48 & 63.33\\
        ST-MEM \cite{stmem} & \xmark &  & \cellcolor{gray!30}76.12 & \textbf{84.50} & \cellcolor{gray!30}62.27 & \cellcolor{gray!30}75.19 & \cellcolor{gray!30}73.05 & \cellcolor{gray!30}64.66 \\
        \midrule
        \textbf{MERL (Ours)} & \cmark & \textbf{0\%} & \textbf{88.21} & \cellcolor{gray!30}78.01 & \textbf{76.77} & \textbf{76.56} & \textbf{74.15} & \textbf{82.86}\\
        \bottomrule
    \end{tabular}
    }
    \vspace{-5pt}
    \label{tab: shift cls}
\end{table*}

While our MERL highlights its zero-shot classification capability, assessing ECG representations via uni-modal tasks after pretraining is more common.
We select linear probing for our evaluation protocol because of its standardized procedures for eSSL methods \cite{clocs, crt, astcl, stmem}.

Tab. \ref{tab:linear-cls} presents the results of linear probing for MERL and existing eSSL methods.
Our MERL consistently outperform eSSL methods across a 1\%-100\% of training data ratio and six datasets.
Notably, MERL's performance with just 1\% data in linear probing on the PTBXL-Super dataset surpasses those of all eSSL methods using 100\% data.
Furthermore, MERL with 10\% data in linear probing outperforms all eSSL methods with 100\% data on the remaining five datasets.
This demonstrates that clinical report supervision enables MERL, to learn discriminative ECG representations with richer semantics.
Interestingly, the second-highest performance on four subsets of the PTBXL with 100\% data linear probing is achieved by a randomly initialized ResNet18, rather than by any eSSL methods.
We speculate this is due to two reasons: (1) for contrastive eSSL approaches, the quality of the learned representation decreases because the positive/negative pairs, generated through naive signal-level augmentation, introduce semantic distortion; and (2) for generative eSSL methods, the ECG representation learned through reconstruction tasks lacks discriminative and high-level semantic information \cite{liu2023improving, liu2023pixmim, he2022masked}.

\subsection{Robustness to Distribution Shift}

Distribution shift refers to scenarios where the test set's ECGs come from a different distribution (often caused by different data sources) than the training set. Among them we focus on the most common distribution shift in healthcare data: domain shift (covariate shift), where the label space are shared but input distributions vary. To evaluate the generalizability and robustness of the learned ECG representation across different sources, we conduct linear probing with eSSL methods and zero-shot MERL under domain shifts: training on one dataset (the `source domain') and testing on another (the `target domain'), which has categories in common with the source domain. 


We prepare the target domain similarly to CLIP \cite{clip}. The details of preparation can be found in the \textit{Appendix}. After preparing the target domain samples, we compare zero-shot MERL with all eSSL methods using 100\% data for linear probing across six target domains. The results are outlined in Tab. \ref{tab: shift cls}. Remarkably, zero-shot MERL 
outperforms all eSSL methods that are linear probed with 100\% data, except in the \textit{PTBXL-Super$\rightarrow$CSN} setting.
We also observe that ST-MEM \cite{stmem} achieves the second-highest overall results. This may be attributed to ST-MEM being pre-trained on a reconstruction task without involving naive signal level data augmentation for constructing positive/negative pairs. This behavior supports our postulation that naive data augmentation in eSSL could impair the robustness of the learned ECG representation.
Overall, these findings underscore that the ECG features learned via MERL are both representative and robust.

\section{Ablation Studies}
\label{sec: ablation}
In this section, we perform comprehensive ablation studies on the key components/design choices, and report the average performance of zero-shot classification and linear probing with 1\% data across six ECG classification datasets.

\noindent\textbf{Loss Function.}
Tab. \ref{tab: abla loss} shows that training with the combination of $\mathcal{L}_{\textrm{CMA}}$ and $\mathcal{L}_{\textrm{UMA}}$ improves performance compared to solely using CMA. 
This suggests that UMA enhances the model's ability to learn ECG representation in the latent space, benefiting downstream tasks.

\begin{table}[ht!]
    \centering
\vspace{-15pt}
\caption{Ablating Loss Function.}
    \scalebox{0.7}{
    \begin{tabular}{cc|cc}
    \toprule[1.2pt]
      $\mathcal{L}_{\textrm{CMA}}$ & $\mathcal{L}_{\textrm{UMA}}$ & Zero-shot & Linear Probing (1\%) \\
    \midrule[1.2pt]
     \cmark   &  & 60.84$\pm$3.8 & 64.25$\pm$2.6 \\
     \cmark & \cmark   & \textbf{61.67$\pm$4.2} & \textbf{65.96$\pm$2.1} \\
    \bottomrule[1.2pt]
    \end{tabular}
}
\vspace{-10pt}
\label{tab: abla loss}
\end{table}

\noindent\textbf{Text Encoder.}
Tab. \ref{tab: abla lm} shows the effects of various text encoders. 
Med-CPT \cite{jin2023medcpt} \footnote{https://huggingface.co/ncbi/MedCPT-Query-Encoder} achieves the highest performance in both tasks. 
The other two text encoders yield suboptimal outcomes. 
We attribute Med-CPT's superior performance to its discriminative and representative text embeddings:
Med-CPT is pre-trained on a text contrastive learning task
 \footnote{This task is implemented on the query-article pair from PubMed search log, and has no overlap with the pre-training dataset in this work.},
differing from other encoders that are pre-trained on masked language modeling tasks. 

\begin{table}[ht!]
    \centering
    \vspace{-15pt}
    \captionof{table}{Effects of Text Encoder Choices.}
    \scalebox{0.65}{
    \begin{tabular}{c|cc}
    \toprule[1.2pt]
     & Zero-shot & Linear Probing (1\%) \\
    \midrule[1.2pt]
    BioClinicalBERT \cite{alsentzer2019publicly} & 72.36$\pm$4.5 & 63.81$\pm$1.8  \\
    \midrule
    PubMedBERT \cite{gu2021domain}& 71.84$\pm$3.2 & 64.29$\pm$2.5  \\
    \midrule
    Med-CPT \cite{jin2023medcpt}& \textbf{75.24$\pm$1.7} & \textbf{65.96$\pm$2.1} \\
    \bottomrule[1.2pt]
    \end{tabular}
    }
    
    \label{tab: abla lm}
    \vspace{-10pt}
\end{table}

\noindent\textbf{Clinical Knowledge Database.}
We further explore the effects of web and local clinical knowledge databases in Tab. \ref{tab: abla database}. 
Eliminating either database results in decreased performance. 
Specifically, removing the web database, SNOMEDCT 
, leads to a notable reduction in performance, attributable to its larger scale compared to the local database. 
This underscores that both clinical knowledge databases are beneficial for zero-shot classification, with the larger-scale database providing more improvements.
\begin{table}[ht!]
    \centering
    \vspace{-15pt}
    \captionof{table}{Benefits of Clinical Knowledge Database}
    \scalebox{0.7}{
    \begin{tabular}{c|c}
    \toprule[1.2pt]
    Database & Zero-shot  \\
    \midrule[1.2pt]
    w/o SNOMEDCT (web) & 72.17$\pm$2.3  \\
    \midrule
    w/o  SCP Statement (local) & 73.62$\pm$1.9   \\
    \midrule
    Ours & \textbf{75.24$\pm$1.7} \\
    \bottomrule[1.2pt]
    \end{tabular}
    }
    \label{tab: abla database}
    \vspace{-10pt}
\end{table}

\noindent\textbf{Data Augmentation Strategies.}
We implement four augmentation strategies and report the results in Tab. \ref{tab: abla ecg aug}. As shown in Fig. \ref{fig: frame} (a), naive data augmentation on ECG signal domain distorts semantic information and reduces the quality of the representation.
Instead, the proposed latent space augmentation demonstrates superior performance compared to other strategies applied to raw ECG signals. 

\begin{table}[ht!]
    \centering
    \vspace{-15pt}
    \captionof{table}{Effect of Diverse ECG Augmentation Strategies.}
    \scalebox{0.7}{
    \begin{tabular}{c|cc}
    \toprule[1.2pt]
    Augmentation & Zero-shot & Linear Probing (1\%) \\
    \midrule[1.2pt]
    Cutout & 73.24$\pm$3.2 & 62.24$\pm$2.7  \\
    \midrule
    Drop & 72.79$\pm$2.5 & 61.14$\pm$2.2  \\
    \midrule
    Gaussian noise & 72.61$\pm$2.7 & 64.17$\pm$1.6  \\
    \midrule
    Latent Dropout (Ours) & \textbf{75.24$\pm$1.7} & \textbf{65.96$\pm$2.1} \\
    \bottomrule[1.2pt]
    \end{tabular}
    }
    
    \label{tab: abla ecg aug}
    \vspace{-10pt}
\end{table}

\noindent\textbf{Dropout Ratio.}
Finally, we implement latent augmentation using various dropout ratios and report the results in Tab. \ref{tab: abla drop ratio}. 
A dropout ratio of 0.1 yields the best results, while both higher and lower ratios lead to decreased performance. 
Therefore, we opt for using a dropout ratio of 0.1 for latent augmentation in our method.

\begin{table}[ht!]
    \centering
    \vspace{-15pt}
    \captionof{table}{Effect of Difference Dropout Ratios}
    \scalebox{0.7}{
    \begin{tabular}{c|cc}
    \toprule[1.2pt]
    Dropout Ratio & Zero-shot & Linear Probing (1\%) \\
    \midrule[1.2pt]
    0.05 & 74.79$\pm$1.4 & 65.23$\pm$1.8  \\
    \midrule
    0.1 & \textbf{75.24$\pm$1.7} & \textbf{65.96$\pm$2.1} \\
    \midrule
    0.15 & 74.53$\pm$2.6 & 64.19$\pm$1.7 \\
    \midrule
    0.2 & 74.25$\pm$2.1 & 64.43$\pm$2.5 \\
    \bottomrule[1.2pt]
    \end{tabular}
    }
    \label{tab: abla drop ratio}
\end{table}

\noindent\textbf{Feature Extractors for ECG.}
Tab. \ref{tab: abla ecg encoder} outlines the ablation study for two ECG feature extractor networks (\textit{i.e.,} backbones): the CNN-based ResNet18\cite{resnet} and the transformer-based ViT-Tiny\cite{dosovitskiy2020image}. 
Results show that performance of MERL with CNN backbone surpasses that of ViT, suggesting CNN are better suited for capturing ECG patterns. The degraded efficacy of the transformer backbone may stem from its tokenization strategy, which discretizes continuous signals, possibly leading to information loss.

\begin{table}[ht!]
    \centering
    \vspace{-15pt}
    \captionof{table}{Effects of ECG Feature Extractor Choices}
    \scalebox{0.7}{
    \begin{tabular}{c|cc}
    \toprule[1.2pt]
    Augmentation & Zero-shot & Linear Probing (1\%) \\
    \midrule[1.2pt]
    ViT\cite{dosovitskiy2020image} & 73.54$\pm$2.3 & 63.53$\pm$2.6  \\
    \midrule
    ResNet\cite{resnet} & \textbf{75.24$\pm$1.7} & \textbf{65.96$\pm$2.1} \\
    \bottomrule[1.2pt]
    \end{tabular}
    }
    
    \label{tab: abla ecg encoder}
    \vspace{-10pt}
\end{table}

\begin{figure}[ht!]
    \centering
    \vspace{-0pt}
    \includegraphics[width=0.99\linewidth]{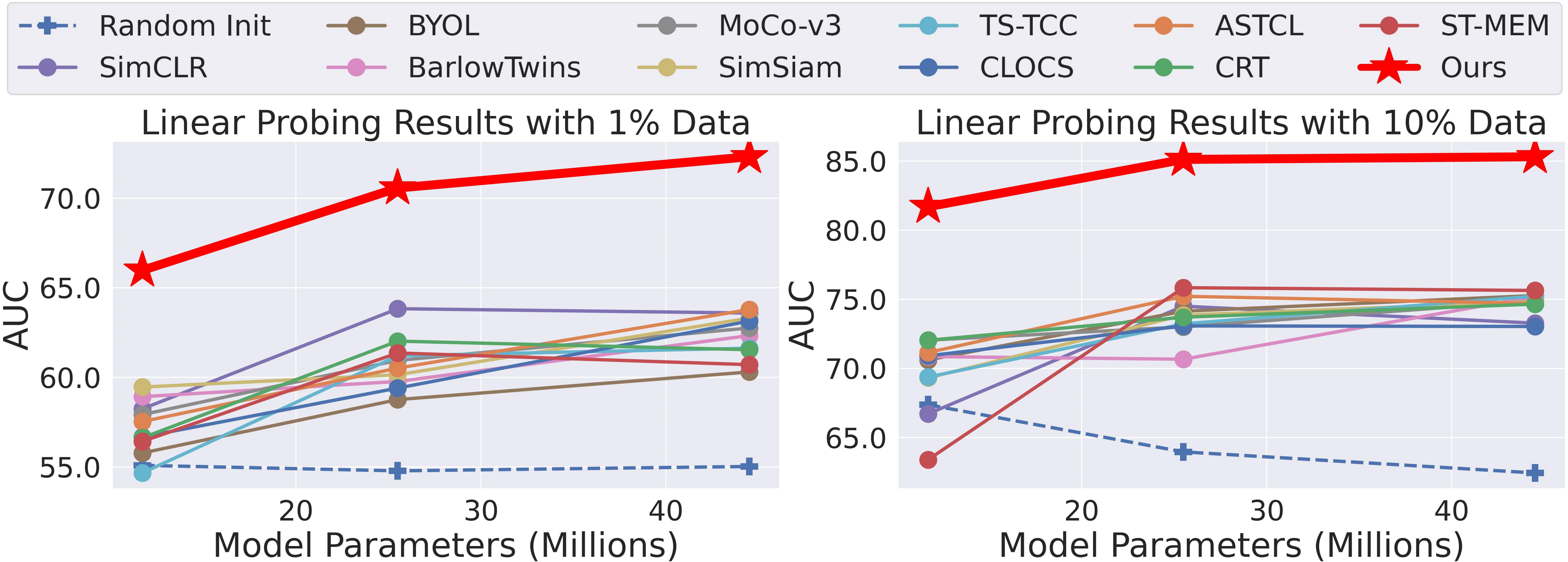}
    \vspace{-4mm}
    \caption{Average linear probing performance on six datasets of \textcolor{red}{MERL} and other eSSL methods with scaled ECG backbones. For ST-MEM, a transformer-based method, we use ViT-Tiny, ViT-Small, and ViT-Base.}
    \label{fig: scale}
    \vspace{-10pt}
\end{figure}

\noindent\textbf{Scalability.} We scale up the backbone size using three models, ResNet18, ResNet50, and ResNet101, for MERL and other eSSL methods, and report their linear probing performance with 1-10\% of data across six datasets in Fig. \ref{fig: scale}. 
Across these scaled models, MERL consistently outperforms the other eSSL methods. 
Significantly, even the smallest ECG backbone in MERL exceeds the performance of the largest backbones in other eSSL methods. 
This highlight the value of clinical reports for representation learning for ECG. 
Moreover, as the backbone size increases, MERL's performance improves, whereas eSSL methods experience varied performance fluctuation with scaled ECG encoder sizes. 
This findings demonstrate that MERL is a scalable, effective, and data-efficient strategy for ECG representation learning.

\section{Conclusion}
We introduce MERL, a scalable and effective multimodal ECG learning framework that incorporates CMA and UMA alignment strategies during training, and CKEPE, a strategy for customizing prompts, during testing.
CKEPE leverages the capabilities of LLMs to extract and restructure clinical knowledge from a provided database, boosting zero-shot MERL to outperform eSSL with linear probing in classification tasks. 
Additionally, we establish the first benchmark that includes 10 eSSL methods and MERL, all pre-trained on the largest public ECG datasets and evaluated across a broad range of classification tasks.
MERL's superior performance in both zero-shot and linear probing tasks underscores the advantages of multimodal ECG learning with report supervision over eSSL methods that only learn representations in the signal domain. 
We hope that both MERL and this benchmark will benefit the research community in ECG representation learning.

\clearpage
\section*{Broader Impact}
Our MERL framework significantly advances ECG classification by utilizing a zero-shot approach that eliminates the need for annotated data in downstream ECG tasks. It outperforms linearly probed self-supervised learning methods through innovative prompt engineering. To support future research, we build the first comprehensive ECG representation learning benchmark, which covers six datasets and introduces domain transfer scenarios.
Our research primarily uses LLMs to generate descriptive prompts. Although we compel the LLM to extract and reformat knowledge from clinical expert-verified databases, the generation process remains not fully controlled and transparent. Therefore, the field of safe, controllable, and trusted generation in clinical applications recognizes the need for further development and exploration.
In the future, we aim to align ECG records with diverse medical data modalities, such as electronic health records, cardiac imaging, and cardiologist reports, to enhance multimodal medical data understanding.

\bibliography{main}
\bibliographystyle{icml2023}

\newpage
\appendix
\onecolumn
\section{Downstream Task Details}
\subsection{Downstream Task Data Split}
We detail the data split in Tab. \ref{tab:split}. For the four subsets of PTBXL, we adhere to the official split from the original work of \cite{wagner2020ptb}. For CPSC2018 \cite{cpsc2018} and CSN \cite{csn1,csn2}, we randomly split the data, and all data split information will be released post-acceptance.

\begin{table}[ht!]
\centering
\caption{Details on Data Split.}
\scalebox{0.8}{
\begin{tabular}{lcrrr}
\toprule[1.2pt]
Dataset & Number of Categories & Train & Valid & Test \\ 
\midrule[1.2pt]
PTBXL-Super \cite{wagner2020ptb} & 5 & 17,084 & 2,146 & 2,158 \\
PTBXL-Sub \cite{wagner2020ptb} & 23 & 17,084 & 2,146 & 2,158 \\
PTBXL-Form \cite{wagner2020ptb} & 19 & 7,197 & 901 & 880 \\
PTBXL-Rhythm \cite{wagner2020ptb} & 12 & 16,832 & 2,100 & 2,098 \\
\midrule
CPSC2018 \cite{cpsc2018} & 9 & 4,950 & 551 & 1,376 \\
CSN \cite{csn1,csn2} & 38 & 16,546 & 1,860 & 4,620 \\
 \bottomrule[1.2pt]
\end{tabular}
}

\label{tab:split}
\end{table}

\subsection{Downstream Task Configuration}

Since not all categories of the target domain are covered by the source domain, we merge target domain categories into the most similar source domain category similarly to CLIP \cite{clip}. For example, `ST wave tilt up' and `ST wave drop down' are merged into `ST-T wave change' when the target domain has a broader range of categories than the source domain. If the target domain includes categories not present in the source domain (e.g., `Wolf-Parkinson-White syndrome' from the CSN dataset), we remove these distinct categories and their associated samples from the target domain. The category relations for the data distribution transfer scenario can be found in Tab. \ref{tab: ptbsuper2cpsc2018}, \ref{tab: ptbsuper2csn}, and\ref{tab: cpsc2018tocsn}. 
We also show hyperparameters for all downstream tasks are listed in Tab. \ref{tab: hyper}.

\begin{table}[ht!]
    \centering
    \caption{Hyperparameter settings on downstream tasks.}
    \scalebox{0.8}{
    \begin{tabular}{ccccccc}
    \toprule[1.2pt]
    & PTBXL-Super & PTBXL-Sub & PTBXL-Form & PTBXL-Rhythm & CPSC2018 & CSN\\
    \midrule[1.2pt]
    Learning rate & 0.001 & 0.001 & 0.001 & 0.001 & 0.001 & 0.001\\
    Batch size & 16 & 16 & 16 & 16 & 16 & 16 \\
    Epochs & 100 & 100 & 100 & 100 & 100 & 100 \\
    Optimizer & AdamW & AdamW & AdamW & AdamW & AdamW & AdamW \\
    Learing rate scheduler & Cosine anealing & Cosine anealing & Cosine anealing & Cosine anealing & Cosine anealing & Cosine anealing \\
    Warump steps & 5 & 5 & 5 & 5 & 5 & 5 \\
    \bottomrule[1.2pt]
    \end{tabular}
    }
    
    \label{tab: hyper}
\end{table}

\begin{table}[ht!]
    \centering
    \caption{Domain transfer category matching for PTBXL-Super to CPSC2018, `None' indicates that there is no category from target domain belongs source domain.}
    \scalebox{0.9}{
    \begin{tabular}{c|c}
    \toprule[1.2pt]
    Source Domain & Target Domain\\
    \midrule[1.2pt]
    HYP & None \\
    NORM & NORM \\
    CD & 1AVB, CRBBB, CLBBB \\
    MI & None \\
    STTC & STE, STD \\
    \bottomrule[1.2pt]
    \end{tabular}
    }
    
    \label{tab: ptbsuper2cpsc2018}
\end{table}

\begin{table}[ht!]
    \centering
    \caption{Domain transfer category matching for PTBXL-Super to CSN, `None' indicates that there is no category from target domain belongs source domain.}
    \scalebox{0.9}{
    \begin{tabular}{c|c}
    \toprule[1.2pt]
    Source Domain & Target Domain\\
    \midrule[1.2pt]
    HYP & RVH, LVH \\
    NORM & SR \\
    CD & 2AVB, 2AVB1, 1AVB, AVB, LBBB, RBBB, STDD \\
    MI & MI \\
    STTC & STTC, STE, TWO, STTU, QTIE, TWC \\
    \bottomrule[1.2pt]
    \end{tabular}
    }
    
    \label{tab: ptbsuper2csn}
\end{table}

\begin{table}[ht!]
    \centering
    \caption{Domain transfer category matching for CPSC2018 to CSN.}
    \scalebox{0.9}{
    \begin{tabular}{c|c}
    \toprule[1.2pt]
    Source Domain & Target Domain\\
    \midrule[1.2pt]
    AFIB & AFIB \\
    VPC & VPB \\
    NORM & SR \\
    1AVB & 1AVB \\
    CRBBB & RBBB \\
    STE & STE \\
    PAC & APB \\
    CLBBB & LBBB \\
    STD & STE, STTC, STTU, STDD \\
    \bottomrule[1.2pt]
    \end{tabular}
    }
    
    \label{tab: cpsc2018tocsn}
\end{table}

\section{Visualization on Learned ECG Representation}

\begin{figure}[t!]
    \centering
    \includegraphics[width=0.99\linewidth]{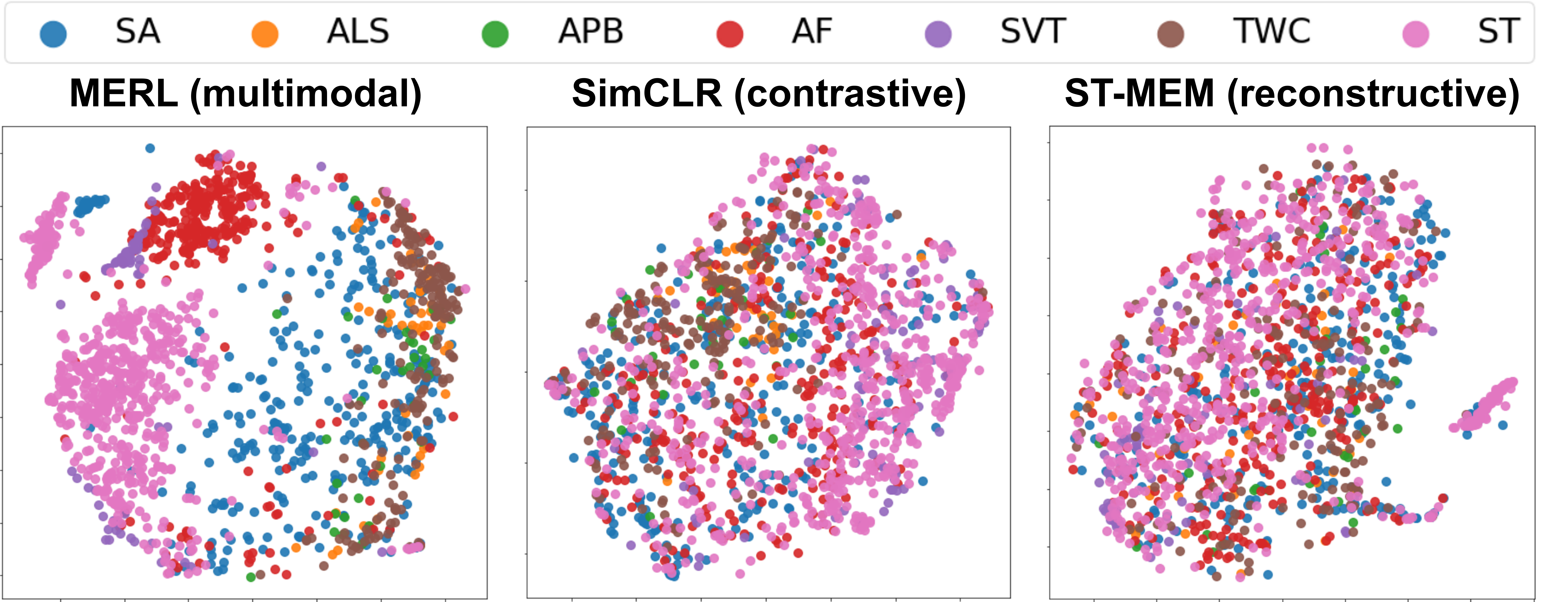}
    \caption{
    The t-SNE visualization of the learned ECG representation after pre-training. We utilize the test set of the CSN dataset, which includes only samples with unique categories and remove categories that have fewer than 50 samples for better visualization.
    }
    \label{fig: tsne vis}
\end{figure}

To further investigate the learned ECG representation, we visualize the last layer output of the ECG encoder using three methods: MERL (multimodal), SimCLR (contrastive), and ST-MEM (reconstructive) on the CSN test set. As Fig. \ref{fig: tsne vis} shows, 
MERL distinguishes the samples with different categories even without supervised learning, while both SimCLR and ST-MEM struggle with mixed ECG features from unique categories, even though the pre-training target of SimCLR aims to learn the distinctiveness of each sample.
This demonstrates that clinical report supervision benefits the ECG encoder in learning more discriminative ECG features. 
Additionally, the visualization in Fig. \ref{fig: tsne vis} indicates the flaw of the reconstructive method in learning high-level discriminative semantics from ECG, because the pre-training target focuses only on low-level signal patterns (e.g., signal intensity). It also highlights the flaw of the contrastive method, which learns representation from semantically distorted samples, as shown in Fig. \ref{fig: frame} (a).

\end{document}